\documentclass[prb,aps,superbib,citeautoscript,twocolumn]{revtex4}
\usepackage{amsmath,bm}
\usepackage{amstext}
\usepackage{epsfig}
\usepackage{xcolor}
\definecolor{dgreen}{rgb}{0,.5,0}
\definecolor{dred}{rgb}{.7,.0,.0}

\begin{document}

\title{Rigorous formulation of two-parameter double-hybrid density-functionals
}

\author{Emmanuel Fromager}
\affiliation{\it 
~\\
Laboratoire de Chimie Quantique,\\
Institut de Chimie, CNRS / Universit\'{e} de Strasbourg,\\
4 rue Blaise Pascal, 67000 Strasbourg, France.\\
}


\begin{abstract}
A two-parameter extension of the density-scaled double hybrid approach
of Sharkas {\it et al.} [J. Chem. Phys. {\bf
134}, 064113 (2011)] is presented. It is based on the explicit treatment
of a fraction of
multideterminantal exact exchange. The connection with
conventional double hybrids is made when neglecting density scaling in
the correlation functional as well as second-order corrections to the density. 
In this context, the fraction $a_{\rm c}$ of second-order M\o ller-Plesset (MP2) 
correlation energy is not necessarily equal to the square of the
fraction $a_{\rm x}$ of
Hartree-Fock exchange. More specifically, it is shown   
that $a_{\rm c}\leq a^2_{\rm x}$, a condition
that conventional semi-empirical double hybrids actually fulfill. 
In addition, a new procedure for
calculating the orbitals, which has a better justification than the one routinely
used, is proposed. Referred to as $\lambda_1$ variant, the corresponding
double hybrid approximation has been tested on a small set
consisting of H$_2$, N$_2$, Be$_2$, Mg$_2$ and Ar$_2$. Three
conventional double hybrids (B2-PLYP, B2GP-PLYP and PBE0-DH) have been
considered. Potential curves obtained with $\lambda_1$- and regular
double hybrids can, in some cases, differ significantly. In particular,
for the weakly bound dimers, the $\lambda_1$ variants bind
systematically more than the regular ones, which is an improvement in
many but not all cases.
Including  
density scaling in
the correlation functionals may of course change the results
significantly. Moreover, optimized effective potentials (OEPs) based on a
partially-interacting system could also be used to generate proper orbitals. Work
is currently in progress in those directions.

\end{abstract}

\maketitle

\section{Introduction}\label{intro}

Sharkas {\it et
al.}~\cite{2blehybrids_Julien} recently proposed a rigorous formulation
of one-parameter double hybrid density-functionals, which is based on
the combination of second-order M\o ller-Plesset (MP2) perturbation
theory with a density-scaled correlation functional. The connection with
regular double hybrids~\cite{2blehybrids_Grimme} could be made when
neglecting density scaling. Note that, in their approach, the fraction
$a_{\rm c}$ of MP2 correlation energy equals the square of the fraction $a_{\rm x}$ of Hartree-Fock (HF)
exchange. Such a condition is actually not fulfilled by conventional
double hybrids. The $a_{\rm c}$ parameter is either chosen irrespective
of $a_{\rm x}$ or expressed in terms of $a_{\rm x}$ but {\it not as} $a_{\rm
c}= a^2_{\rm x}$. In
the latter case, one can for example refer to the
Perdew-Burke-Ernzerhof-zero double hybrid (PBE0-DH) functional of Br\'{e}mond
and Adamo~\cite{pbe0-dh_Adamo}, which is characterized by $a_{\rm c}=
a^3_{\rm x}$.\\

This work deals with the rigorous formulation of density-scaled
two-parameter double hybrids. It is organized as follows: an {\it exact}
two-parameter energy expression is firstly derived in Sec.~\ref{mdXsection}.
Approximate formulations are then investigated for defining single
hybrid (Sec.~\ref{secHF-OEP}) and double hybrid (Sec.~\ref{secMP2-OEP}) energy expressions.
In Sec.~\ref{convDH_sec}, the connection with conventional double
hybrids is made. In addition, a new procedure for
computing orbitals is proposed, defining thus what is referred to as
$\lambda_1$ variant of the double
hybrids. A summary of the different approximations that have been
formulated is then given in Sec.~\ref{summary-sec}. 
Following the computational details
(Sec.~\ref{comp-details-sec}), results obtained with regular and
$\lambda_1$-double hybrids on a small test set consisting of H$_2$,
N$_2$, Be$_2$, Mg$_2$ and Ar$_2$ are presented and discussed in Sec.~\ref{results-sec}.

\section{Theory}\label{theory}

\subsection{Multideterminantal exact exchange}\label{mdXsection}
The approach recently proposed by Sharkas {\it et
al.}~\cite{2blehybrids_Julien} for deriving rigorous 
one-parameter double-hybrid functionals 
is based on the separation of the universal
Hohenberg-Kohn~\cite{hktheo} functional
$F[n]=F^{\lambda_1}[n]+\overline{E}^{\lambda_1}_{\rm Hxc}[n]
$
into a partially-interacting contribution 
\begin{eqnarray}\label{HKFm1def}\begin{array} {l}
{\displaystyle
F^{\lambda_1}[n]=\underset{\Psi\rightarrow n}{\rm min}\langle \Psi\vert
\hat{T}+\lambda_1\hat{W}_{ee}\vert\Psi\rangle}\\
\\
\hspace{1.2cm}=\langle \Psi^{\lambda_1}[n]\vert
\hat{T}+\lambda_1\hat{W}_{ee}\vert\Psi^{\lambda_1}[n]\rangle,
\end{array}
\end{eqnarray}
with $0\leq\lambda_1\leq 1$, and the {\it complement} $\lambda_1$-dependent
Hartree-exchange-correlation (Hxc) density-functional   





\begin{eqnarray}\label{Hxcm1decomp}\begin{array} {l}
\overline{E}^{\lambda_1}_{\rm Hxc}[n]=(1-\lambda_1)E_{\rm Hx}[n]
+\overline{E}^{\lambda_1}_{\rm c}[n],\\
\\
\overline{E}^{\lambda_1}_{\rm c}[n]=E_{\rm c}[n]-E^{\lambda_1}_{\rm
c}[n],
\end{array}
\end{eqnarray}
where $E_{\rm Hx}[n]=\langle \Phi^{\rm KS}[n]\vert \hat{W}_{ee}\vert
\Phi^{\rm KS}[n]\rangle$ is the usual exact Hartree-exchange (Hx) term based on the
non-interacting Kohn-Sham (KS) determinant~\cite{kstheo}. The correlation functionals
$E_{\rm c}[n]$ and $E^{\lambda_1}_{\rm c}[n]$ correspond to fully and
partially $\lambda_1$-interacting systems, respectively.
The exact ground-state energy is then expressed as follows, according to
the variational principle\cite{hktheo}, 
\begin{eqnarray}\label{energymindensm1}\begin{array} {l}
{\displaystyle
E = \underset{n}{\rm min} \left\{ \langle \Psi^{\lambda_1}[n]\vert
\hat{T}+\lambda_1\hat{W}_{ee}+\hat{V}_{\rm
ne}\vert\Psi^{\lambda_1}[n]\rangle\right.}\\
\left. \hspace{3.25cm} + \overline{E}^{\lambda_1}_{\rm Hxc}[n]\right\},
\end{array}
\end{eqnarray}
where $\hat{V}_{\rm ne}=\int d\mathbf{r}\; v_{\rm ne}(\mathbf{r})
\,\hat{n}(\mathbf{r})$ is the nuclear potential operator. Rewritten in terms
of a minimization over wave functions, Eq.~(\ref{energymindensm1}) becomes
\begin{eqnarray}\label{energyminpsi}\begin{array} {l}
{\displaystyle
E = \underset{\Psi}{\rm min}\left\{ \langle \Psi\vert
\hat{T}+\lambda_1\hat{W}_{ee}+\hat{V}_{\rm ne}\vert\Psi\rangle
+\overline{E}^{\lambda_1}_{\rm Hxc}[n_{\Psi}]\right\}  
}
\\
\\
\hspace{0.4cm}  =  \langle \Psi^{\lambda_1}\vert
\hat{T}+\lambda_1\hat{W}_{ee}+\hat{V}_{\rm ne}\vert\Psi^{\lambda_1}\rangle
+\overline{E}^{\lambda_1}_{\rm Hxc}[n_{\Psi^{\lambda_1}}],
\end{array}
\end{eqnarray}
where $\Psi^{\lambda_1}$ fulfills the self-consistent equation
\begin{eqnarray}\label{selfconsisteqlambda1}\begin{array} {l}
{\displaystyle
\left(\hat{T}+\lambda_1\hat{W}_{ee}+\hat{V}_{\rm ne}+
\hat{\overline{V}}^{\lambda_1}_{\rm
Hxc}[n_{\Psi^{\lambda_1}}]\right)\vert \Psi^{\lambda_1}\rangle
=\mathcal{E}^{\lambda_1}\vert\Psi^{\lambda_1}\rangle,
}
\\
\\
{\displaystyle
\hat{\overline{V}}^{\lambda_1}_{\rm Hxc}[n]=\int d\mathbf{r}\;
\;\frac{\delta \overline{E}^{\lambda_1}_{\rm Hxc}}{\delta
n(\mathbf{r})}[n]\,\hat{n}(\mathbf{r}).
}
\end{array}
\end{eqnarray}
In order to introduce a second scaling factor $\lambda_2$, and thus
derive two-parameter double hybrids, let us consider the following partitioning of
the {\it complement} $\lambda_1$-Hxc functional:  
\begin{equation}\label{mdXCcm1def}
\overline{E}^{\lambda_1}_{\rm Hxc}[n]= \overline{E}^{\lambda_1}_{\rm
Hx,md}[n]+\overline{E}^{\lambda_1}_{\rm c,md}[n],
\end{equation}
which is based on the multideterminantal definition of the exact
exchange ({md}EXX), as introduced by Toulouse, Gori-Giorgi and Savin
~\cite{TousrXmd,PaolasrXmd,PaolasrXmd_prb} in the context of
range-separated density-functional theory (DFT), 
\begin{equation}\label{mdXcm1def}
\overline{E}^{\lambda_1}_{\rm Hx,md}[n]=(1-\lambda_1)\langle \Psi^{\lambda_1}[n]\vert \hat{W}_{ee} 
\vert \Psi^{\lambda_1}[n]\rangle.
\end{equation}
The corresponding {\it complement} correlation functional differs therefore
from the one given in Eq.~(\ref{Hxcm1decomp}): according to
Eqs. (\ref{mdXCcm1def}) and (\ref{mdXcm1def}) it can be expressed as 
\begin{eqnarray}\label{mdccm1expression}\begin{array} {l}
{\displaystyle
\overline{E}^{\lambda_1}_{\rm c,md}[n]=\overline{E}^{\lambda_1}_{\rm
c}[n]-(1-\lambda_1)\Delta_{\rm c}^{\lambda_1}[n],
}
\\
\\
{\displaystyle
\Delta_{\rm c}^{\lambda_1}[n]=\left( \langle
\Psi^{\lambda_1}[n]\vert \hat{W}_{ee} \vert \Psi^{\lambda_1}[n]\rangle
\right.}\\ 
\hspace{3.6cm}\left.-\langle \Phi^{\rm KS}[n]\vert \hat{W}_{ee}\vert
\Phi^{\rm KS}[n]\rangle\right).\\
\end{array}
\end{eqnarray}
Recombined with Eq.~({\ref{energymindensm1}}), Eq.~(\ref{mdXcm1def}) leads
to the explicit calculation of 100 \% of {md}EXX. In the spirit of usual hybrid
functionals, we want to keep the flexibility of treating only a fraction
$\lambda_2$ ($0\leq \lambda_2\leq 1$) of {md}EXX explicitly. For that
purpose, we split the {\it
complement} {md}Hx functional as follows:
\begin{eqnarray}\label{mdXcm1m2}\begin{array} {l}
\overline{E}^{\lambda_1}_{\rm Hx,md}[n]=
(\lambda_2-\lambda_1)\langle
\Psi^{\lambda_1}[n]\vert \hat{W}_{ee}\vert
\Psi^{\lambda_1}[n]\rangle
\\
\\
\hspace{1.9cm}+(1-\lambda_2)\langle \Psi^{\lambda_1}[n]\vert
\hat{W}_{ee} \vert \Psi^{\lambda_1}[n]\rangle,\\
\end{array}   
\end{eqnarray}
and introduce the {\it complement} $\lambda_1$- and
$\lambda_2$-dependent Hxc density-functional energy: 
\begin{eqnarray}\label{mdHXccm1cm2def}\begin{array} {l}
\overline{E}^{\lambda_1,\lambda_2}_{\rm
Hxc}[n]=
(1-\lambda_2)\langle \Psi^{\lambda_1}[n]\vert
\hat{W}_{ee} \vert \Psi^{\lambda_1}[n]\rangle+\overline{E}^{\lambda_1}_{\rm c,md}[n]\\
\\
\hspace{1.5cm}=-(\lambda_2-\lambda_1)\langle \Psi^{\lambda_1}[n]\vert
\hat{W}_{ee}\vert
\Psi^{\lambda_1}[n]\rangle\\
\\
\hspace{1.9cm}+\overline{E}^{\lambda_1}_{\rm
Hx,md}[n]+\overline{E}^{\lambda_1}_{\rm c,md}[n],

\end{array}
\end{eqnarray}
which, according to Eqs.~(\ref{mdXCcm1def}) and (\ref{mdccm1expression}), can
be rewritten as 
\begin{eqnarray}\label{mdXcm1m2BIS}\begin{array} {l}
\overline{E}^{\lambda_1,\lambda_2}_{\rm
Hxc}[n]=
-(\lambda_2-\lambda_1)\langle \Psi^{\lambda_1}[n]\vert
\hat{W}_{ee}\vert
\Psi^{\lambda_1}[n]\rangle
+\overline{E}^{\lambda_1}_{\rm Hxc}[n]
\\
\\
\hspace{1.5cm}=(1-\lambda_2)E_{\rm Hx}[n] +
\overline{E}^{\lambda_1}_{\rm c}[n]
\\
\\
\hspace{4.25cm}
+(\lambda_1-\lambda_2)\Delta_{\rm
c}^{\lambda_1}[n].\\

\end{array}   
\end{eqnarray}
Using the following expressions based on a uniform coordinate scaling of the
density~\cite{2blehybrids_Julien,PRB_Levy_Perdew_Eclambda}:  
\begin{eqnarray}\label{scaled_corr_fun}\begin{array} {l}
E^{\lambda_1}_{\rm c}[n]=\lambda^2_1E_{\rm
c}[n_{1/\lambda_1}],\\
\\
{\displaystyle
\Delta_{\rm c}^{\lambda_1}[n]=\left .\frac{\partial E^{\lambda}_{\rm
c}[n]}{\partial \lambda}\right|_{\lambda=\lambda_1},
}\\
\\
n_{1/\lambda_1}(\mathbf{r})=(1/\lambda_1)^3n(\mathbf{r}/\lambda_1),
\end{array}
\end{eqnarray}
as well as Eq.~(\ref{Hxcm1decomp}), the more explicit {\it density-scaled two-parameter
} (DS2) form
\begin{eqnarray}\label{HXccm1cm2final}\begin{array} {l}
\overline{E}^{\lambda_1,\lambda_2}_{\rm
Hxc}[n]=(1-\lambda_2)E_{\rm Hx}[n] +E_{\rm c}[n]-\lambda^2_1E_{\rm c}[n_{1/\lambda_1}]\\
\\
\hspace{1.85cm}+2\lambda_1(\lambda_1-\lambda_2)E_{\rm c}[n_{1/\lambda_1}]\\
\\
\hspace{1.85cm}+\lambda_1^2(\lambda_1-\lambda_2){\displaystyle\left . \frac{\partial E_{\rm
c}[n_{1/\lambda}]}{\partial\lambda }\right|_{\lambda=\lambda_1}},
\\
\end{array}
\end{eqnarray}
is obtained and the {\it exact} ground-state energy in Eq.~(\ref{energymindensm1}) can
then be rewritten as follows: 
\begin{eqnarray}\label{energymindensm1m2}\begin{array} {l}
{\displaystyle
E = \underset{n}{\rm min} \left\{ \langle \Psi^{\lambda_1}[n]\vert
\hat{T}+\lambda_2\hat{W}_{ee}+\hat{V}_{\rm
ne}\vert\Psi^{\lambda_1}[n]\rangle\right.}\\
\left. \hspace{3.15cm} + \overline{E}^{\lambda_1,\lambda_2}_{\rm
Hxc}[n]\right\},
\end{array}
\end{eqnarray}\\
or, in terms of minimization over local potentials,
\begin{eqnarray}\label{energyminVm1m2}\begin{array} {l}
{\displaystyle
E = \underset{v}{\rm min} \left\{ \langle \Psi^{\lambda_1}[v]\vert
\hat{T}+\lambda_2\hat{W}_{ee}+\hat{V}_{\rm
ne}\vert\Psi^{\lambda_1}[v]\rangle\right.}\\
\left. \hspace{3.15cm} + \overline{E}^{\lambda_1,\lambda_2}_{\rm
Hxc}[n_{\Psi^{\lambda_1}[v]}]\right\},
\end{array}
\end{eqnarray}
where $\Psi^{\lambda_1}[v]$ denotes the ground state of the
$\lambda_1$-interacting Hamiltonian $\hat{T}+\lambda_1\hat{W}_{ee}+\int
d\mathbf{r}\,v(\mathbf{r})\,\hat{n}(\mathbf{r})$. 
It is important to notice that, for a fixed $\lambda_1$ value, the {\it
exact} minimizing potential $v^{\lambda_1}$ is the one which ensures that the density of
$\Psi^{\lambda_1}[v^{\lambda_1}]$ equals the {\it exact} ground-state density.
Therefore, it {\it does not} depend on $\lambda_2$. When choosing
$\lambda_2=\lambda_1$, the minimization over potentials in
Eq.~(\ref{energyminVm1m2}) can be replaced by a minimization over wave
functions so that Eq.~(\ref{energyminpsi}) is recovered and, according to
Eq.~(\ref{selfconsisteqlambda1}), the {\it exact} potential
$v^{\lambda_1}$ can be expressed as 
\begin{equation}\label{Xvlam1}
v^{\lambda_1}(\mathbf{r})=v_{\rm
ne}(\mathbf{r})+{\displaystyle \frac{\delta{\overline{E}^{\lambda_1}_{\rm Hxc}}}{\delta
n(\mathbf{r})}[n_{\Psi^{\lambda_1}}]}.
\end{equation}
As a result, for any $\lambda_1$ and $\lambda_2$ values, the {\it exact} ground-state energy can be written as 
\begin{eqnarray}\label{exactenergycm1cm2}\begin{array} {l}
{\displaystyle
E = 
\langle \Psi^{\lambda_1}\vert
\hat{T}+\lambda_2\hat{W}_{ee}+\hat{V}_{\rm ne}\vert\Psi^{\lambda_1}\rangle
+\overline{E}^{\lambda_1,\lambda_2}_{\rm Hxc}[n_{\Psi^{\lambda_1}}].

}
\\
\end{array}
\end{eqnarray}
If $\lambda_2\neq\lambda_1$, the energy cannot be obtained
straightfowardly from a minimization over wave functions: 
\begin{eqnarray}\label{energyminPsim1m2}\begin{array} {l}
{\displaystyle
E \neq \underset{\Psi}{\rm min} \left\{ \langle \Psi\vert
\hat{T}+\lambda_2\hat{W}_{ee}+\hat{V}_{\rm
ne}\vert\Psi\rangle\right.}\\
\left. \hspace{2.45cm} + \overline{E}^{\lambda_1,\lambda_2}_{\rm
Hxc}[n_{\Psi}]\right\},
\end{array}
\end{eqnarray}
since the minimizing wave function would be eigenfunction of a
$\lambda_2$-interacting system and therefore could not be equal to
$\Psi^{\lambda_1}$, which is eigenfunction of a $\lambda_1$-interacting
system. As discussed in Sec.~\ref{secHF-OEP}, it is in principle possible to adapt regular optimized effective potential
(OEP) methods to this context in order to implement
Eq.~(\ref{energyminVm1m2}).\\

\subsection{Density-scaled two-parameter single hybrids}\label{secHF-OEP}

This section deals with the formulation of single hybrid functionals
based on the energy expression given in Eq.~(\ref{energyminVm1m2}). The
latter can be considered as
a generalization of the standard OEP approach, which is based on a non-interacting
KS system,
since $\Psi^{\lambda_1}[v]$ is the ground state of a
partially-interacting system. In principle, any wave-function-theory-based model
can thus be combined rigorously with OEPs. We consider in this work the
optimization of the potential using the HF approximation
$\Phi^{\lambda_1}[v]$ to the exact wave function $\Psi^{\lambda_1}[v]$.
Following the terminology of Sharkas {\it et
al.}~\cite{2blehybrids_Julien}, we refer to this
approach as DS2-HF-OEP approximation. In this case, the energy becomes 
\begin{eqnarray}\label{energyminVHFlambda12}\begin{array} {l}
E_{\mbox{\tiny HF-OEP}}^{\mbox{\tiny DS2},\lambda_1,\lambda_2}=\underset{v}{\rm min}
\;E_{\mbox{\tiny HF-OEP}}^{\mbox{\tiny DS2},\lambda_1,\lambda_2}[v],\\ 
\\
{\displaystyle
E_{\mbox{\tiny HF-OEP}}^{\mbox{\tiny DS2},\lambda_1,\lambda_2}[v] = \left\{ \langle
\Phi^{\lambda_1}[v]\vert
\hat{T}+\lambda_2\hat{W}_{ee}+\hat{V}_{\rm
ne}\vert\Phi^{\lambda_1}[v]\rangle\right.}\\
\\
\left.\hspace{4.10cm}+\overline{E}^{\lambda_1,\lambda_2}_{{\rm Hxc
}}[n_{\Phi^{\lambda_1}[v]}]\right\}.
\\
\\
\end{array}
\end{eqnarray}
Such an approach can be interpreted as an exchange-only-type OEP
calculation based on a $\lambda_1$-interacting system. As a result, the
DS2-HF-OEP energy gradient has no simple analytical
expression, like in the KS-OEP scheme~\cite{Yang_prl_oep}. However, it can be efficiently computed 
from a HF-type linear response vector, using a decomposition of the
potential in an auxiliary basis of gaussian
functions. Work is currently in progress in this
direction. If we simply denote $\Phi$ and $n$ the converged DS2-HF-OEP
determinant and its density, the energy can be reexpressed, according to
Eq.~(\ref{HXccm1cm2final}), as  
\begin{eqnarray}\label{DS2HFOEPener_likeDH}\begin{array} {l}
\\
E_{\mbox{\tiny HF-OEP}}^{\mbox{\tiny DS2},\lambda_1,\lambda_2}= \langle
\Phi\vert
\hat{T}+\hat{V}_{\rm
ne}\vert\Phi\rangle
+E_{\rm H}[n]+\lambda_2E^{\mbox{\tiny HF}}_{\rm x}[\Phi]\\
\\
+(1-\lambda_2)E_{\rm x}[n]
+E_{\rm  c}[n]-\lambda_1(2\lambda_2-\lambda_1)E_{\rm
c}[n_{1/\lambda_1}]
\\
\\
+\lambda_1^2(\lambda_1-\lambda_2){\displaystyle\left . \frac{\partial E_{\rm c}[n_{1/\lambda}]}{\partial\lambda }\right|_{\lambda=\lambda_1}}
,
\end{array}
\end{eqnarray}
where $E^{\mbox{\tiny HF}}_{\rm x}[\Phi]$ is the HF exchange
energy. We thus obtain a {\it density-scaled two-parameter 
hybrid} (DS2H) exchange-correlation energy:  
\begin{eqnarray}\label{DS2HFOEPXCener_likeSH}\begin{array} {l}
\\
E_{\rm xc,\mbox{\tiny DS2H}}^{\lambda_1,\lambda_2}= 
\lambda_2E^{\mbox{\tiny HF}}_{\rm x}[\Phi]
+(1-\lambda_2)E_{\rm x}[n]
+E_{\rm  c}[n]\\
\\
\hspace{1.8cm}-\lambda_1(2\lambda_2-\lambda_1)E_{\rm
c}[n_{1/\lambda_1}]
\\
\\
\hspace{1.8cm}+\lambda_1^2(\lambda_1-\lambda_2){\displaystyle\left .
\frac{\partial E_{\rm c}[n_{1/\lambda}]}{\partial\lambda
}\right|_{\lambda=\lambda_1}}.
\end{array}
\end{eqnarray}
Note that, in the particular case $\lambda_1=\lambda_2$, the {\it exact}
DS2-HF-OEP potential
can be expressed in terms of a functional derivative like in the exact
theory (see Sec.~\ref{mdXsection}). Indeed, the DS2-Hxc functional in
Eq.~(\ref{HXccm1cm2final}) reduces then to a DS1-Hxc functional which
corresponds to the {\it complement}
$\lambda_1$-Hxc functional:
\begin{eqnarray}\label{DS1hxcfun}\begin{array} {l}
\overline{E}^{\lambda_1,\lambda_1}_{\rm
Hxc}[n]=(1-\lambda_1)E_{\rm Hx}[n] +E_{\rm c}[n]-\lambda^2_1E_{\rm
c}[n_{1/\lambda_1}]\\
\\
\hspace{1.5cm}=\overline{E}^{\lambda_1}_{\rm
Hxc}[n].
\end{array}
\end{eqnarray}
Moreover, using the {\it one-to-one} correspondance between the density $n_
{\Phi}$ of any single determinant $\Phi$ and the local potential
$v^{\lambda_1}[n_{\Phi}]$ such that the density of the HF determinant
${\Phi^{\lambda_1}[v^{\lambda_1}[n_{\Phi}]]}$ equals $n_{\Phi}$, the
energy in Eq.~(\ref{energyminVHFlambda12}) can be simply obtained by minimization over single
determinants, that is without using OEPs. It then corresponds to the
DS1H energy of Sharkas {\it et al.}\cite{2blehybrids_Julien} which
equals       
\begin{eqnarray}\label{DS1H_reduction}\begin{array} {l}
{\displaystyle
E_{\mbox{\tiny DS1H}}^{\lambda_1}= \underset{\Phi}{\rm min} \left\{ \langle \Phi\vert
\hat{T}+\lambda_1\hat{W}_{ee}+\hat{V}_{\rm
ne}\vert\Phi\rangle
+ \overline{E}^{\lambda_1}_{\rm
Hxc}[n_{\Phi}]\right\}.
}
\end{array}
\end{eqnarray}
The minimizing determinant
$\Phi^{\lambda_1}$ fulfills the self-consistent equation
\begin{equation}\label{DS1H_lam1eqlam2}
\left(\hat{T}+\lambda_1\hat{U}_{\rm
HF}[\Phi^{\lambda_1}]+\hat{V}_{\rm ne}+\hat{\overline{V}}^{\lambda_1}_{\rm
Hxc}[n_{\Phi^{\lambda_1}}]
\right)\vert
\Phi^{\lambda_1}\rangle =\mathcal{E}^{\lambda_1}_{\mbox{\tiny DS1H} }\vert
\Phi^{\lambda_1}\rangle,
\end{equation}
where $\hat{U}_{\rm HF}[\Phi^{\lambda_1}]$ is the nonlocal HF
potential, so that, in this particular case, the minimizing potential in    
Eq.~(\ref{energyminVHFlambda12}) can be expressed as 
\begin{equation}\label{oep_ds1h}
v_{\mbox{\tiny DS1H}}^{\lambda_1}(\mathbf{r})=v_{\rm ne}(\mathbf{r})+\frac{\delta
\overline{E}^{\lambda_1}_{\rm Hxc}}{\delta
n(\mathbf{r})}[n_{\Phi^{\lambda_1}}],
\end{equation}
and the exchange-correlation in Eq.~(\ref{DS2HFOEPXCener_likeSH}) reduces to 
\begin{eqnarray}\label{DS1H_sharkas}\begin{array} {l}
\\
E_{\rm xc,\mbox{\tiny DS1H}}^{\lambda_1}= 
\lambda_1E^{\mbox{\tiny HF}}_{\rm x}[\Phi]
+(1-\lambda_1)E_{\rm x}[n]
+E_{\rm  c}[n]\\
\\
\hspace{1.7cm}-\lambda_1^2E_{\rm
c}[n_{1/\lambda_1}].
\end{array}
\end{eqnarray}

\subsection{Density-scaled two-parameter double hybrids}\label{secMP2-OEP}

This section deals with the formulation of double hybrid functionals
based on the energy expression given in Eq.~(\ref{energyminVm1m2}). In
the following, we denote $v^{\lambda_1}_0$ {\it any approximation} to the
exact minimizing potential $v^{\lambda_1}$. An approximate expression
for the energy is thus obtained:
\begin{eqnarray}\label{energyV0m1m2}\begin{array} {l}
{\displaystyle
E_0^{\lambda_1,\lambda_2}=\langle \Psi^{\lambda_1}_0\vert
\hat{T}+\lambda_2\hat{W}_{ee}+\hat{V}_{\rm
ne}\vert\Psi^{\lambda_1}_0\rangle}
+ \overline{E}^{\lambda_1,\lambda_2}_{\rm
Hxc}[n_{\Psi^{\lambda_1}_0}],
\end{array}
\end{eqnarray}
where $\Psi^{\lambda_1}_0$ is the ground state of the
$\lambda_1$-interacting system defined by $v^{\lambda_1}_0$: 
\begin{equation}\label{eigenvalueeqm1V}
\left(\hat{T}+\lambda_1\hat{W}_{ee}+\hat{V}_0^{\lambda_1}\right)\vert
\Psi^{\lambda_1}_0\rangle
=\mathcal{E}_0^{\lambda_1}\vert
\Psi^{\lambda_1}_0\rangle,
\end{equation}
with $\hat{V}_0^{\lambda_1}=\int d\mathbf{r}\;
v^{\lambda_1}_0(\mathbf{r})\,\hat{n}(\mathbf{r})$.
The energy in
Eq.~(\ref{energyV0m1m2}) can be rewritten as 
\begin{eqnarray}\label{energyV0m1m2forMP2}\begin{array} {l}
{\displaystyle
E_0^{\lambda_1,\lambda_2} =\mathcal{E}^{\lambda_1}_0+(\lambda_2-\lambda_1)\langle
\Psi^{\lambda_1}_0\vert
\hat{W}_{ee}
\vert\Psi^{\lambda_1}_0\rangle}\\
\\
\hspace{2.1cm}+
\langle\Psi^{\lambda_1}_0\vert\hat{V}_{\rm
ne}-\hat{V}^{\lambda_1}_0\vert\Psi^{\lambda_1}_0\rangle
\\
\\
\hspace{2.10cm} 
+\overline{E}^{\lambda_1,\lambda_2}_{\rm
Hxc}[n_{\Psi^{\lambda_1}_0}],
\end{array}
\end{eqnarray}
which is convenient when applying, as we propose in the following, MP perturbation theory
to the $\lambda_1$-interacting system described by
Eq.~(\ref{eigenvalueeqm1V}). The zeroth-order wave function is chosen to
be the determinant $\Phi^{\lambda_1}_0$ which fulfills the HF-type
equation 
\begin{equation}\label{HFeq_2blehyb}
\left(\hat{T}+\lambda_1\hat{U}_{\rm
HF}[\Phi^{\lambda_1}_0]+\hat{V}^{\lambda_1}_0\right)\vert
\Phi^{\lambda_1}_0\rangle =\mathcal{E}^{\lambda_1}_{\rm HF}\vert
\Phi^{\lambda_1}_0\rangle.
\end{equation}
Note that the Brillouin theorem is fulfilled in this context, which
means that the wave function contains only double excitations through
first order and, as a result, the density remains unchanged through
first order~\cite{pra_MBPTn-srdft}. The perturbation expansion of the fictitious
$\lambda_1$-interacting energy through second order equals: 
\begin{equation}\label{fictitiousenerMP2}
\mathcal{E}^{\lambda_1}_0=
\langle \Phi^{\lambda_1}_0\vert
\hat{T}+\lambda_1\hat{W}_{ee}+\hat{V}^{\lambda_1}_0\vert
\Phi^{\lambda_1}_0\rangle
+\lambda^2_1E^{(2)}_{\rm MP}+\ldots,
\end{equation}
where $E^{(2)}_{\rm MP}$ is the conventional MP2 energy correction
calculated with the $\lambda_1$-interacting orbitals and orbital
energies obtained from Eq.~(\ref{HFeq_2blehyb}). Using
the simplified perturbation expansion for the second term in the right-hand side of
Eq.~(\ref{energyV0m1m2forMP2}): 
\begin{eqnarray}\label{2ndterm}\begin{array} {l}
{\displaystyle
\langle \Psi^{\lambda_1}_0\vert
\hat{W}_{ee}
\vert\Psi^{\lambda_1}_0\rangle=
\langle \Phi^{\lambda_1}_0\vert
\hat{W}_{ee}
\vert\Phi^{\lambda_1}_0\rangle}+2\lambda_1E^{(2)}_{\rm MP} +\ldots,
\\
\\
\end{array}
\end{eqnarray}
we obtain the final energy expression through second order:
\begin{eqnarray}\label{2blehybenerexpB}\begin{array} {l}
{\displaystyle
E_0^{
\lambda_1,\lambda_2}=
\langle \Phi^{\lambda_1}_0\vert
\hat{T}+\lambda_2\hat{W}_{ee}+\hat{V}_{\rm ne}\vert
\Phi^{\lambda_1}_0\rangle}\\
\\
+\overline{E}^{\lambda_1,\lambda_2}_{\rm
Hxc}[n_{\Phi^{\lambda_1}_0}]
+\lambda_1(2\lambda_2-\lambda_1)E^{(2)}_{\rm
MP}\\
\\
{\displaystyle
+\int d\mathbf{r}\;\left(v_{\rm
ne}(\mathbf{r})-v_0^{\lambda_1}(\mathbf{r})+\frac{\delta
\overline{E}^{\lambda_1,\lambda_2}_{\rm Hxc}}{\delta
n(\mathbf{r})}[n_{\Phi^{\lambda_1}_0}]\right)\delta
n^{(2)}(\mathbf{r})
}
\\
+\ldots,
\\
\\
\end{array}
\end{eqnarray}
where $\delta n^{(2)}(\mathbf{r})$ denotes the
second-order correction to the density associated to $\Psi^{\lambda_1}_0$. If we simply denote $\Phi$ and
$n$ the zeroth-order determinant $\Phi^{\lambda_1}_0$ and its density,
respectively, the exchange-correlation energy has, according to
Eq.~(\ref{HXccm1cm2final}), the form of a {\it density scaled
two-parameter double hybrid} (DS2DH) functional:
\begin{eqnarray}\label{DS2DHXC_likeDH}\begin{array} {l}
\\
E_{\rm xc,\mbox{\tiny DS2DH}}^{\lambda_1,\lambda_2}= 
\lambda_2E^{\mbox{\tiny HF}}_{\rm x}[\Phi]
+(1-\lambda_2)E_{\rm x}[n]
+E_{\rm  c}[n]\\
\\
\hspace{1.8cm}-\lambda_1(2\lambda_2-\lambda_1)E_{\rm
c}[n_{1/\lambda_1}]
\\
\\
\hspace{1.8cm}+\lambda_1^2(\lambda_1-\lambda_2){\displaystyle\left .
\frac{\partial E_{\rm c}[n_{1/\lambda}]}{\partial\lambda
}\right|_{\lambda=\lambda_1}}
\\
\\
\hspace{1.8cm}+\lambda_1(2\lambda_2-\lambda_1)E^{(2)}_{\rm MP}
\\
\\
{\displaystyle
+\int d\mathbf{r}\;\left(v_{\rm
ne}(\mathbf{r})-v_0^{\lambda_1}(\mathbf{r})+\frac{\delta
\overline{E}^{\lambda_1,\lambda_2}_{\rm Hxc}}{\delta
n(\mathbf{r})}[n]\right)\delta
n^{(2)}(\mathbf{r}).
}
\end{array}
\end{eqnarray}
Let us consider the particular
case $\lambda_1=\lambda_2$. According to Eqs.~(\ref{DS1hxcfun}) and (\ref{oep_ds1h}), when
choosing the DS1H determinant $\Phi^{\lambda_1}$ and potential $v_{\mbox{\tiny
DS1H}}^{\lambda_1}$
as $\Phi$ and $v_0^{\lambda_1}$, the second-order-density-correction
term in the right-hand side of Eq.~(\ref{DS2DHXC_likeDH}) cancels out
and the DS1DH exchange-correlation energy of Sharkas {\it et
al.}~\cite{2blehybrids_Julien} is
recovered:  
\begin{eqnarray}\label{DS1DHXC_sharkas}\begin{array} {l}
\\
E_{\rm xc,\mbox{\tiny DS1DH}}^{\lambda_1}= 
\lambda_1E^{\mbox{\tiny HF}}_{\rm x}[\Phi]
+(1-\lambda_1)E_{\rm x}[n]
+E_{\rm  c}[n]\\
\\
\hspace{1.8cm}-\lambda_1^2 E_{\rm
c}[n_{1/\lambda_1}]
+\lambda_1^2E^{(2)}_{\rm MP}.
\end{array}
\end{eqnarray}
As shown in Sec.~\ref{convDH_sec}, the DS2DH functional defined in
Eq.~(\ref{DS2DHXC_likeDH}) can be connected with conventional double
hybrids when neglecting both second-order corrections to the
density as well as the density scaling in the correlation functional. 

\subsection{Connection with conventional double hybrids}\label{convDH_sec}

In order to connect regular double hybrids with the one derived in
Sec.~\ref{secMP2-OEP}, we neglect both second-order corrections to the
density as well as the density scaling in the correlation functional:
\begin{eqnarray}\label{approx_for_conv2DHs}\begin{array} {l}
\delta n^{(2)}(\mathbf{r})\approx 0,\;\; E_{\rm c}[n_{1/\lambda}]\approx
E_{\rm c}[n].
\end{array}
\end{eqnarray}
The DS2H and DS2DH exchange-correlation energies in
Eqs.~(\ref{DS2HFOEPXCener_likeSH}) and
(\ref{DS2DHXC_likeDH})
reduce then to {\it two-parameter hybrid} (2H) and {\it two-parameter
double hybrid} (2DH) exchange-correlation energies, respectively:
\begin{eqnarray}\label{2Hxccorr}\begin{array} {l}
E_{\rm {xc},\mbox{\tiny 2H}}^{a_{\rm x},a_{\rm c}}= 
a_{\rm x}E^{\mbox{\tiny HF}}_{\rm x}[\Phi]
+(1-a_{\rm x})E_{\rm x}[n]
\\
\\
\hspace{1.6cm}+(1-a_{\rm c})E_{\rm c}[n],
\end{array}
\end{eqnarray}
and 
\begin{eqnarray}\label{xcener_likeDH}\begin{array} {l}
E_{\rm {xc},\mbox{\tiny 2DH}}^{a_{\rm x},a_{\rm c}}= 
a_{\rm x}E^{\mbox{\tiny HF}}_{\rm x}[\Phi]
+(1-a_{\rm x})E_{\rm x}[n]\\
\\
\hspace{1.6cm}+(1-a_{\rm c})E_{\rm c}[n]+a_{\rm c}E^{(2)}_{\rm MP},
\end{array}
\end{eqnarray}
where we have introduced the two parameters $a_{\rm x}$ and $a_{\rm c}$ defined as
follows:   
\begin{eqnarray}\label{axac}\begin{array} {l}
a_{\rm x}=\lambda_2,
\\
\\
a_{\rm c}=\lambda_1(2\lambda_2-\lambda_1)\\
\\
\hspace{0.45cm}   = a_{\rm x}^2-(a_{\rm x}-\lambda_1)^2.
\\
\end{array}
\end{eqnarray}


Let us first notice that, by contrast to KS
second-order perturbation theory~\cite{kspt2_Yang} (KS-PT2), single
excitations do not appear in the 2DH energy
expression. Indeed, the latter originates from a MP-type calculation,
where the Brillouin theorem therefore applies, and which is based on a
partially $\lambda_1$-interacting system. In addition, it is readily seen from
Eq.~(\ref{axac}) that, in this context, $a_{\rm c}\leq
a^2_{\rm x}$. Interestingly, the various conventional double hybrids considered in
Table~\ref{lambda1values} fulfill this condition.
Note that the {\it one-parameter double hybrid} (1DH) approximation~\cite{2blehybrids_Julien} is recovered when $\lambda_1=\lambda_2$,
or, in terms of $a_{\rm x}$ and $a_{\rm c}$, $a_{\rm c}=a_{\rm x}^2$.
For given values of $\lambda_1$ and $\lambda_2$, a unique
set of $a_{\rm x}$ and $a_{\rm c}$ parameters can be defined according
to Eq.~(\ref{axac}). Reciprocally, for given $a_{\rm x}$ and $a_{\rm c}$
values, we can define the two scaling factors
$\lambda_1$ and $\lambda_2$ as follows:
\begin{eqnarray}\label{lambda12_functionaxac}\begin{array} {l}
\lambda_2=a_{\rm x},
\\
\\
\lambda_1=a_{\rm x}\pm \sqrt{a^2_{\rm x}-a_{\rm c}}.
\\
\end{array}
\end{eqnarray}
When $a_{\rm c}=0$, the correlation energy is fully described by the
correlation functional which means that the fictitious
$\lambda_1$-interacting system should be the non-interacting (KS)
one, that is $\lambda_1=0$. The standard exchange-only-type KS-OEP scheme with $a_{\rm x}$ as fraction of exact
exchange is thus recovered. We therefore conclude 
\begin{eqnarray}\label{lambda12_functionaxacfinal}\begin{array} {l}
\lambda_1=a_{\rm x}-\sqrt{a^2_{\rm x}-a_{\rm c}}.
\\
\end{array}
\end{eqnarray}
Interestingly, the {\it linearly scaled one-parameter double hybrid} functional (LS1DH) derived recently by Toulouse {\it et
al.}~\cite{2bleHybrids_lambda3_Julien}, which is characterized by
$a_{\rm c}=a^3_{\rm x}$, is recovered here, though density scaling is
neglected,   
when $\lambda_1$ equals
\begin{eqnarray}\label{lambda1LS1DH}\begin{array} {l}
\lambda^{\mbox{\tiny LS1DH}}_1=a_{\rm x}(1-\sqrt{1-a_{\rm x}}).
\\
\end{array}
\end{eqnarray}
\\
Let us now focus on the calculation of the orbitals. As pointed out in
Sec.~\ref{secMP2-OEP}, the DS2DH and thus the 2DH exchange-correlation
energies are based on HF-type orbitals calculated for a
$\lambda_1$-interacting system. A natural choice of orbitals would
therefore be the DS2-HF-OEP ones. Since density scaling is neglected, we
will refer to them as $\lambda_1$-OEP-2H orbitals and the corresponding
2DH energy will be referred to as $\lambda_1$-OEP-2DH. The
$\lambda_1$-OEP-2H scheme,
which is an OEP-type calculation, can be formulated as follows,
according to Eqs.~(\ref{HXccm1cm2final}), (\ref{energyminVHFlambda12}) and
(\ref{approx_for_conv2DHs}) and (\ref{axac}):
\begin{eqnarray}\label{2HFOEPapproxAXAC}\begin{array} {l}
E_{\mbox{\tiny $\lambda_1$-OEP-2H}}^{a_{\rm x},a_{\rm c}}=\underset{v}{\rm min}
\;E_{\mbox{\tiny $\lambda_1$-OEP-2H}}^{a_{\rm x},a_{\rm c}}[v],\\ 
\\
{\displaystyle
E_{\mbox{\tiny $\lambda_1$-OEP-2H}}^{a_{\rm x},a_{\rm c}}[v] = \Big\{ \langle
\Phi^{\lambda_1}[v]\vert
\hat{T}+\hat{V}_{\rm
ne}\vert\Phi^{\lambda_1}[v]\rangle}
\\
\\
\hspace{1.0cm} +E_{\rm H}[n_{\Phi^{\lambda_1}[v]}]+a_{\rm x}E^{\mbox{\tiny HF}}_{\rm
x}[\Phi^{\lambda_1}[v]]\\
\\
\hspace{1.0cm}+(1-a_{\rm x})E_{\rm x}[n_{\Phi^{\lambda_1}[v]}]
+(1-a_{\rm c})E_{\rm
c}[n_{\Phi^{\lambda_1}[v]}]\Big\},
\end{array}
\end{eqnarray}
where $\Phi^{\lambda_1}[v]$ fulfills the $\lambda_1$-interacting HF
equation 
\begin{equation}\label{lam1HFeq}
\left(\hat{T}+\lambda_1\hat{U}_{\rm
HF}[\Phi^{\lambda_1}[v]]+\hat{V}\right)\vert
\Phi^{\lambda_1}[v]\rangle =\mathcal{E}^{\lambda_1}_{\rm HF}[v]\vert
\Phi^{\lambda_1}[v]\rangle,
\end{equation}
with 
$ \hat{V}=\int
d\mathbf{r}\,v(\mathbf{r})\,\hat{n}(\mathbf{r})$ and $\lambda_1$ defined
in Eq.~(\ref{lambda12_functionaxacfinal}). On the hand, for conventional
double hybrids, the orbitals are obtained from a 2H calculation which can 
be formulated as~\cite{2blehybrids_Grimme}    
\begin{eqnarray}\label{regular2SHAXAC}\begin{array} {l}
E_{\mbox{\tiny 2H}}^{a_{\rm x},a_{\rm c}}=\underset{\Phi}{\rm min}
{\displaystyle
\Big\{ \langle
\Phi\vert
\hat{T}+\hat{V}_{\rm
ne}\vert\Phi\rangle}
+E_{\rm H}[n_{\Phi}]+a_{\rm x}E^{\mbox{\tiny HF}}_{\rm
x}[\Phi]\\
\\
\hspace{2.3cm}+(1-a_{\rm x})E_{\rm x}[n_{\Phi}]
+(1-a_{\rm c})E_{\rm
c}[n_{\Phi}]\Big\}.
\end{array}
\end{eqnarray}
The minimizing determinant $\tilde{\Phi}$ in Eq.~(\ref{regular2SHAXAC})
fulfills the 2H equation
\begin{eqnarray}\label{hybDFTeq}\begin{array} {l}
\Big(\hat{T}+\hat{V}_{\rm ne}+a_{\rm x}\hat{U}_{\rm
HF}[\tilde{\Phi}]+(1-a_{\rm x})\hat{V}_{\rm Hx}[n_{\tilde{\Phi}}]\\
\\
\hspace{0.55cm}+(1-a_{\rm c})\hat{V}_{\rm c}[n_{\tilde{\Phi}}]\Big)\vert
\tilde{\Phi}\rangle =\mathcal{E}_{\mbox{\tiny 2H}}\vert
\tilde{\Phi}\rangle,
\end{array}
\end{eqnarray}
and, therefore, can be interpreted as the HF determinant associated to a
$\lambda_2$-interacting system since, according to
Eq.~(\ref{lambda12_functionaxac}), $\lambda_2=a_{\rm
x}$. However, as shown in Sec.~\ref{secMP2-OEP}, the 2DH
exchange-correlation functional in Eq.~(\ref{xcener_likeDH}) can be
justified when it is based on HF-type orbitals associated to a $\lambda_1$-interacting
Hamiltonian. This condition ensures that single excitations do
not contribute to the exchange-correlation energy. In fact, the
conventional calculation of the orbitals is simply deduced from
Eq.~(\ref{2HFOEPapproxAXAC}) when replacing the minimization over local
potentials by a minimization over single determinants, which is
justified only when $\lambda_2=\lambda_1$ that is equivalent to
$a_{\rm c}= a_{\rm x}^2$ (see Sec.~\ref{secHF-OEP}). As shown in Table~\ref{lambda1values},
conventional double hybrids do not fulfill the latter condition. In this respect,
double hybrids based on the $\lambda_1$-OEP-2H approximation have a better justification than
the regular ones. Let us finally consider a possible alternative to the
$\lambda_1$-OEP-2H scheme, which would not require the calculation of
an OEP and would have a computational cost similar to
conventional 2H calculations. As shown in
Sec.~\ref{mdXsection}, in the {\it exact} theory and for a fixed
$\lambda_1$ value, the OEP should not
depend on $\lambda_2$. As an approximation, referred to as
$\lambda_1$-2H in the following, we assume that this statement
still holds within the {\it approximate} $\lambda_1$-OEP-2H scheme. As
a result, an approximate potential can be obtained when choosing, in
Eq.~(\ref{energyminVHFlambda12}), $\lambda_2=\lambda_1$ instead of
$\lambda_2=a_{\rm x}$.
In this particular case, according to Eqs.~(\ref{DS1hxcfun}), (\ref{DS1H_lam1eqlam2}), and (\ref{approx_for_conv2DHs}), the $\lambda_1$-2H determinant $\tilde{\Phi}^{\prime}$
fulfills the modified 2H equation     
\begin{eqnarray}\label{lam1hybDFTeq}\begin{array} {l}
\Big(\hat{T}+\hat{V}_{\rm ne}+a^{\prime}_{\rm x}\hat{U}_{\rm
HF}[\tilde{\Phi}^{\prime}]+(1-a^{\prime}_{\rm x})\hat{V}_{\rm
Hx}[n_{\tilde{\Phi}^{\prime}}]\\
\\
\hspace{0.55cm}+(1-a^{\prime}_{\rm c})\hat{V}_{\rm c}[n_{\tilde{\Phi}^{\prime}}]\Big)\vert
\tilde{\Phi}^{\prime}\rangle =\mathcal{E}^{\prime}_{\mbox{\tiny 2H}}\vert
\tilde{\Phi}^{\prime}\rangle,
\end{array}
\end{eqnarray}
where the regular $a_{\rm x}$ and $a_{\rm c}$ parameters have been replaced by
$\lambda_1$ and $\lambda_1^2$, respectively:
\begin{eqnarray}\label{hybDFTeqlambda1}\begin{array} {l}
a_{\rm x}\rightarrow a^{\prime}_{\rm x}=\lambda_1=a_{\rm
x}-\sqrt{a^2_{\rm x}-a_{\rm c}}\\
\\
a_{\rm c}\rightarrow a^{\prime}_{\rm c}=\lambda^2_1=

2a_{\rm x}\left(a_{\rm x}-\sqrt{a^2_{\rm x}-a_{\rm c}}\right)-a_{\rm c}.\\
\end{array}
\end{eqnarray}
We thus ensure that the orbitals are computed from a
$\lambda_1$-interacting system. This procedure   
basically consists in approximating the {\it exact} potential
$v^{\lambda_1}$ by 
\begin{eqnarray}\label{lam1hybDFTeqpot}\begin{array} {l}
{\displaystyle
v_{\mbox{\tiny $\lambda_1$-2H}}(\mathbf{r})=v_{\rm ne}(\mathbf{r})+
(1-a^{\prime}_{\rm x})\frac{\delta{{E}_{\rm Hx}}}{\delta
n(\mathbf{r})}[n_{\tilde{\Phi}^{\prime}}]
}
\\
\\
{\displaystyle
\hspace{2.9cm}+
(1-a^{\prime}_{\rm c})\frac{\delta{{E}_{\rm c}}}{\delta
n(\mathbf{r})}[n_{\tilde{\Phi}^{\prime}}].
}
\end{array}
\end{eqnarray}
Using this potential and $\tilde{\Phi}^{\prime}$
as $v^{\lambda_1}_0$ and $\Phi^{\lambda_1}_0$
in Eq.~(\ref{2blehybenerexpB}), we obtain from
Eqs.~(\ref{approx_for_conv2DHs}) and (\ref{axac}) the following
expressions for the $\lambda_1$-2DH energy: 
\begin{eqnarray}\label{lam1DHener_nodenscal}\begin{array} {l}
E_{\mbox{\tiny
$\lambda_1$-2DH}}^{a_{\rm x},a_{\rm c}}
= 
 \langle
\tilde{\Phi}^{\prime}\vert
\hat{T}+\hat{V}_{\rm
ne}\vert\tilde{\Phi}^{\prime}\rangle
+E_{\rm H}[\tilde{n}^{\prime}]
+a_{\rm x}E^{\mbox{\tiny HF}}_{\rm
x}[\tilde{\Phi}^{\prime}]\\
\\
\hspace{1.6cm}+(1-a_{\rm x})E_{\rm x}[\tilde{n}^{\prime}]\\
\\

\hspace{1.6cm}+(1-a_{\rm c})E_{\rm c}[\tilde{n}^{\prime}]
+a_{\rm c}E^{(2)\prime}_{\rm MP}
\\
\\
 \hspace{1.3cm}=E_{\mbox{\tiny
$\lambda_1$-2H}}^{a_{\rm x},a_{\rm c}}
+a_{\rm c}E^{(2)\prime}_{\rm MP},

\end{array}
\end{eqnarray}
where $\tilde{n}^{\prime}$ denotes the density of
$\tilde{\Phi}^{\prime}$ and $E^{(2)\prime}_{\rm MP}$ is the regular MP2 energy correction
calculated with $\lambda_1$-2H orbitals and orbital energies. 
The $\lambda_1$-2H and $\lambda_1$-2DH exchange-correlation energies are
therefore expressed respectively as
\begin{eqnarray}\label{lam1XCSH_nodenscal}\begin{array} {l}
E_{\rm xc,\mbox{\tiny
$\lambda_1$-2H}}^{a_{\rm x},a_{\rm c}}
=a_{\rm x}E^{\mbox{\tiny HF}}_{\rm
x}[\tilde{\Phi}^{\prime}]
+(1-a_{\rm x})E_{\rm x}[\tilde{n}^{\prime}]\\
\\
\hspace{1.9cm}+(1-a_{\rm c})E_{\rm c}[\tilde{n}^{\prime}],
\end{array}
\end{eqnarray}
and
\begin{eqnarray}\label{lam1XCDH_nodenscal}\begin{array} {l}
E_{\rm xc,\mbox{\tiny
$\lambda_1$-2DH}}^{a_{\rm x},a_{\rm c}}
=a_{\rm x}E^{\mbox{\tiny HF}}_{\rm
x}[\tilde{\Phi}^{\prime}]
+(1-a_{\rm x})E_{\rm x}[\tilde{n}^{\prime}]\\
\\
\hspace{1.9cm}+(1-a_{\rm c})E_{\rm c}[\tilde{n}^{\prime}]+a_{\rm
c}E^{(2)\prime}_{\rm MP}.
\end{array}
\end{eqnarray}
Let us stress that the $\lambda_1$-2H and $\lambda_1$-2DH energies as defined in 
Eq.~(\ref{lam1DHener_nodenscal}), and thus the corresponding
exchange-correlation energies in Eqs.~(\ref{lam1XCSH_nodenscal}) and
(\ref{lam1XCDH_nodenscal}), are
obtained from the same energy expressions as in the regular 2H and 2DH
schemes, using standard $a_{\rm x}$ and $a_{\rm
c}$ values. The difference comes from 
the orbitals which are calculated with the modified $a^{\prime}_{\rm x}$
and $a^{\prime}_{\rm c}$ coefficients.
\subsection{Summary}\label{summary-sec}

A two-parameter extension of the DS1DH scheme proposed recently by Sharkas {\it et
al.}~\cite{2blehybrids_Julien}
has been derived. It is based on the explicit treatment of a fraction of
mdEXX which requires, in the general case, the calculation of an OEP
for a partially-interacting system. Computing this OEP at the HF
level of approximation leads to the DS2-HF-OEP scheme where the energy
has a DS2H form. In addition, it was shown that, using any
approximate potential, DS2DHs can be defined. The connection between the
latter and regular double hybrids is made when neglecting both second-order
corrections to the density and density scaling. In this case, the
DS2-HF-OEP approximation reduces to the $\lambda_1$-OEP-2H one and the
corresponding DS2DH scheme (which is based on the $\lambda_1$-OEP-2H potential) 
is then referred to as $\lambda_1$-OEP-2DH. As an alternative to the
$\lambda_1$-OEP-2H calculation, the $\lambda_1$-2H scheme, where the OEP
calculation is replaced by a 2H one with modified exchange and
correlation coefficients, has been proposed. A $\lambda_1$-2DH
approximation could thus be defined from the $\lambda_1$-OEP-2DH energy expression
using $\lambda_1$-2H orbitals instead of the $\lambda_1$-OEP-2H ones. In this work,
only $\lambda_1$-2H and $\lambda_1$-2DH results will be shown.
The OEP-based schemes, which are currently under implementation, will be presented
in a separate paper.

\section{Computational details}\label{comp-details-sec}

Both $\lambda_1$-2H and $\lambda_1$-2DH energy expressions in
Eq.~(\ref{lam1DHener_nodenscal}) can be simply implemented 
using a
regular DFT code that can perform double hybrid calculations. A  
development version~\cite{Andy_kspt2} of the DALTON
program package~\cite{daltonpack} has been used in this work. 
In a first step, a $\lambda_1$-2H calculation is done. It
consists in performing a 2H calculation with the
modified exchange and correlation coefficients defined in
Eq.~(\ref{hybDFTeqlambda1}). The corresponding orbitals and orbital
energies are then used, in a second step, to compute the scaled MP2
term. In a third step, the $\lambda_1$-2H energy is obtained from a regular one-iteration 2H
calculation, using the $\lambda_1$-2H orbitals as starting
orbitals. Three conventional double hybrids have been considered: the
B2-PLYP~\cite{2blehybrids_Grimme} and B2GP-PLYP~\cite{B2GP_Martin} functionals based on the Becke 88 (B) exchange functional and the
Lee-Yang-Parr (LYP) correlation functional, and the
PBE0-DH~\cite{pbe0-dh_Adamo} functional
which is based on the Perdew-Burke-Ernzerhof (PBE) exchange
functional~\cite{dft-Perdew-PRL1996a} and the Perdew-Wang (PW)
correlation functional~\cite{pw91_XC_fun}. The corresponding exchange
and correlation coefficients are given in Table~\ref{lambda1values}.
Calculations have been performed on a test set consisting of H$_2$,
N$_2$, Be$_2$, Mg$_2$ and Ar$_2$. The following basis sets were used:
cc-pVQZ~\cite{augQZbe} for H$_2$ and N$_2$, aug-cc-pVQZ~\cite{augQZbe}
for Be$_2$ and
Mg$_2$, and aug-cc-pVTZ~\cite{pVTZ_ar} for Ar$_2$.

\section{Results and discussion}\label{results-sec}

In this section we compare, in terms of accuracy, the regular B2-PLYP,
B2GP-PLYP and PBE0-DH double hybrids with their $\lambda_1$ variants,
as defined in Sec.~\ref{convDH_sec}. The potential curves computed for
H$_2$ and N$_2$ (see Fig.~\ref{Potcurves_h2_n2}) show a systematic
lowering of the total energy when using $\lambda_1$-double hybrids instead
of the regular ones. This can be analyzed when decomposing the difference between the
$\lambda_1$-2DH and regular double hybrid energies into a single hybrid term
 $\Delta E_{\rm SH}=E_{\mbox{\tiny $\lambda_1$-2H}}^{a_{\rm
x},a_{\rm c}}-E_{\mbox{\tiny 2H}}^{a_{\rm x},a_{\rm c}}$ (see
Eqs.~(\ref{regular2SHAXAC}) and (\ref{lam1DHener_nodenscal})) and a scaled
MP2 term $a_{\rm c}\Delta E_{\rm
MP2}=a_{\rm c}(E^{(2)\prime}_{\rm MP}-E^{(2)}_{\rm MP})$ (see
Eq.~(\ref{lam1DHener_nodenscal})). As illustrated in Fig.~\ref{HL_gap}
(a), for
H$_2$, the $\lambda_1$-2H energy is always greater than the regular 2H
one ($\Delta E_{\rm SH}\geq 0$) which is due to the fact that the $\lambda_1$-2H orbitals are not
optimized for the regular 2H energy expression (based on $a_{\rm x}$ and
$a_{\rm c}$) but for a modified one (based on $a^{\prime}_{\rm x}$ and
$a^{\prime}_{\rm c}$). This positive difference is then 
compensated when adding the scaled MP2
contribution. According to Eq.~(\ref{hybDFTeqlambda1}), the fraction
$a^{\prime}_{\rm x}$ of
HF exchange used to compute the $\lambda_1$-2H orbitals is lower than
the regular fraction $a_{\rm x}$. This leads to a smaller HOMO-LUMO gap
(see Fig.~\ref{HL_gap} (b))
and thus to a larger scaled MP2 correction. Note, however, that
B2GP-PLYP gives the largest $a_{\rm c}\Delta E_{\rm MP2}$ term in
absolute value while PBE0-DH, for which $a^{\prime}_{\rm x}$ differs the most
from $a_{\rm x}$, has the largest reduction (in absolute value) of HOMO-LUMO
gap. It could be explained by the fact that the scaled MP2 correction is sensitive not only to
the orbital energies but also to the orbitals. Results obtained for N$_2$
(not shown) lead to the same conclusion. Returning to the potential
curves in Fig.~\ref{Potcurves_h2_n2}, we finally note that the $\lambda_1$ variant slightly
improves the accuracy of the double hybrids in the vicinity of the
equilibrium distance.\\
Concerning the weakly-bound systems Be$_2$, Mg$_2$ and Ar$_2$, the
interaction energy curves in Fig.~\ref{Potcurves_vdw} show that regular double
hybrids and
their $\lambda_1$ variants behave similarly at large distances but the
latter bind more than the former. This often leads to slightly more accurate
results unless the regular double hybrid, like PBE0-DH for Be$_2$ and Mg$_2$,
already overbinds. It would of course be interesting to evaluate the
effects of density scaling in the correlation functional. Indeed, in the light
of Sharkas {\it et
al.}~\cite{2blehybrids_Julien} study, results 
are expected to change significantly.\\ 

\section{Conclusions}

A rigorous derivation of two-parameter double hybrids (2DHs) has been
presented. It is based on the combination of the DS1DH scheme of Sharkas {\it et
al.}\cite{2blehybrids_Julien} with the explicit treatment of a fraction of
multideterminantal exact exchange. The connection with regular double
hybrids is made when neglecting both density scaling and second-order
corrections to the density. It then appears, in this context, that the
fraction of second-order M\o ller-Plesset (MP2) energy correlation is smaller or equal to the square of
the fraction of Hartree Fock (HF) exchange. Interestingly, various conventional
semi-empirical double
hybrids fulfill this condition. In the light of those derivations, a new
procedure for calculating the orbitals, which is more justified than the
one used routinely, has been proposed. It still consists in performing a
two-parameter hybrid calculation, but with modified exchange and
correlation coefficients. Preliminary results
presented in this work show that, in such a scheme which is referred to
as $\lambda_1$-2DH, the MP2 energy contribution
is, in absolute value, larger than the regular one. As a result, 
$\lambda_1$-2DH and regular double hybrid potential curves can, in some
cases, differ significantly. In particular,
for the tested weakly bound dimers, the $\lambda_1$ variants bind
systematically more than the regular ones, which is often but not always
an improvement. Including  
density scaling in
the correlation functionals may of course change the results
significantly. 
This still needs to be investigated.
Moreover, optimized effective potentials (OEPs) based on a
partially-interacting system could also be used to generate proper orbitals
. Work
is currently in progress in those directions.
\begin{acknowledgments}
E.F. thanks ANR (contract DYQUMA), Andrew Teale for his help in
computing the $\lambda_1$-2DH energy and his comments on this work, as well as Alexandrina Stoyanova and
Julien Toulouse for fruitful discussions.
\end{acknowledgments}


%
\bibliography{basisset,molecule,srdft,dft,wft}
\clearpage

\textbf{FIGURE CAPTIONS}

\begin{figure}
\caption{\label{Potcurves_h2_n2} Fromager, Journal of Chemical Physics}
\begin{center}
\begin{tabular}{c}
\resizebox{14cm}{!}{\includegraphics{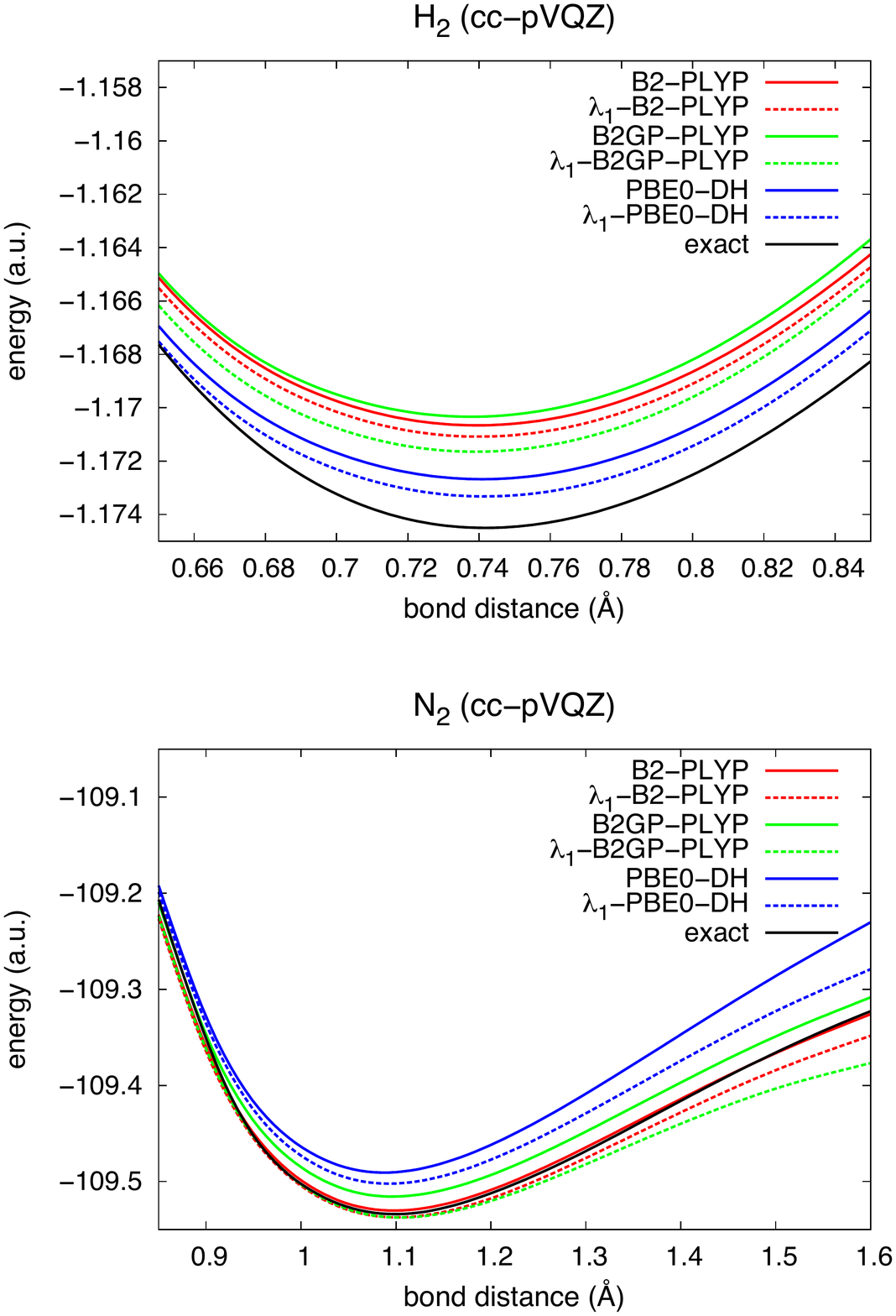}}\\ 
\end{tabular}
\end{center}
\end{figure}

\begin{figure}
\caption{\label{HL_gap} Fromager, Journal of Chemical Physics}
\begin{center}
\begin{tabular}{cc}
\resizebox{18cm}{!}{\includegraphics{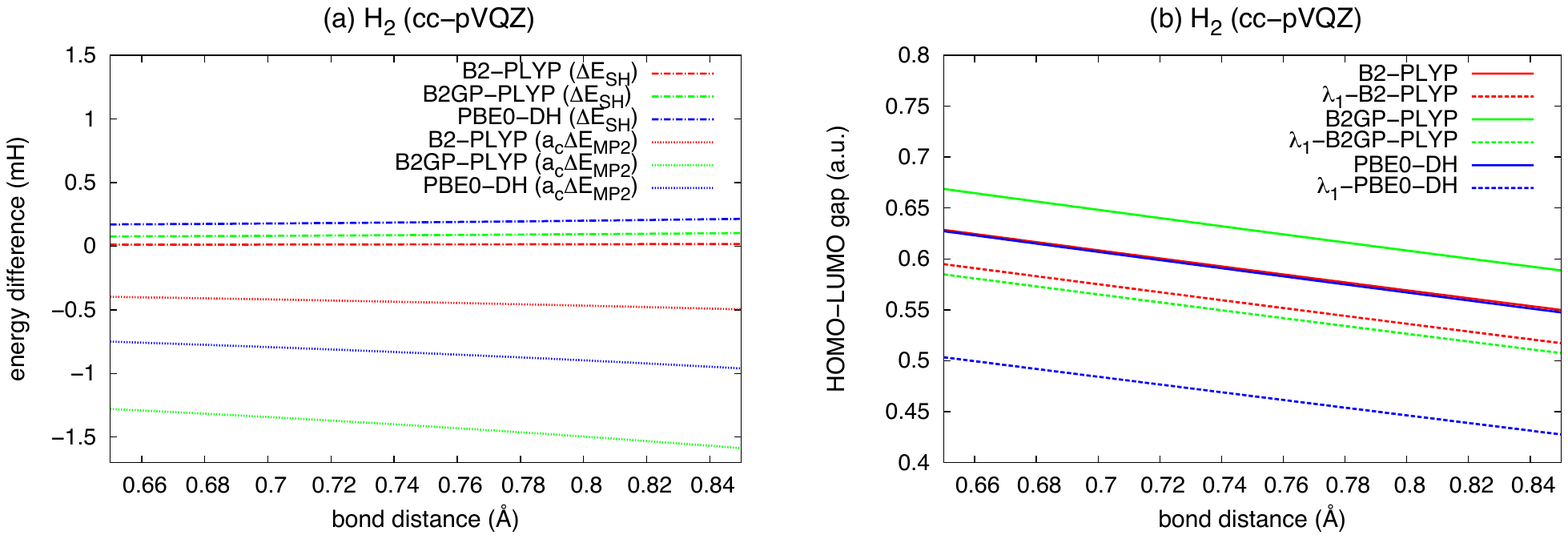}} & 
\end{tabular}
\end{center}
\end{figure}

\begin{figure}
\caption{\label{Potcurves_vdw} Fromager, Journal of Chemical Physics}
\begin{center}
\begin{tabular}{c}
\resizebox{15cm}{!}{\includegraphics{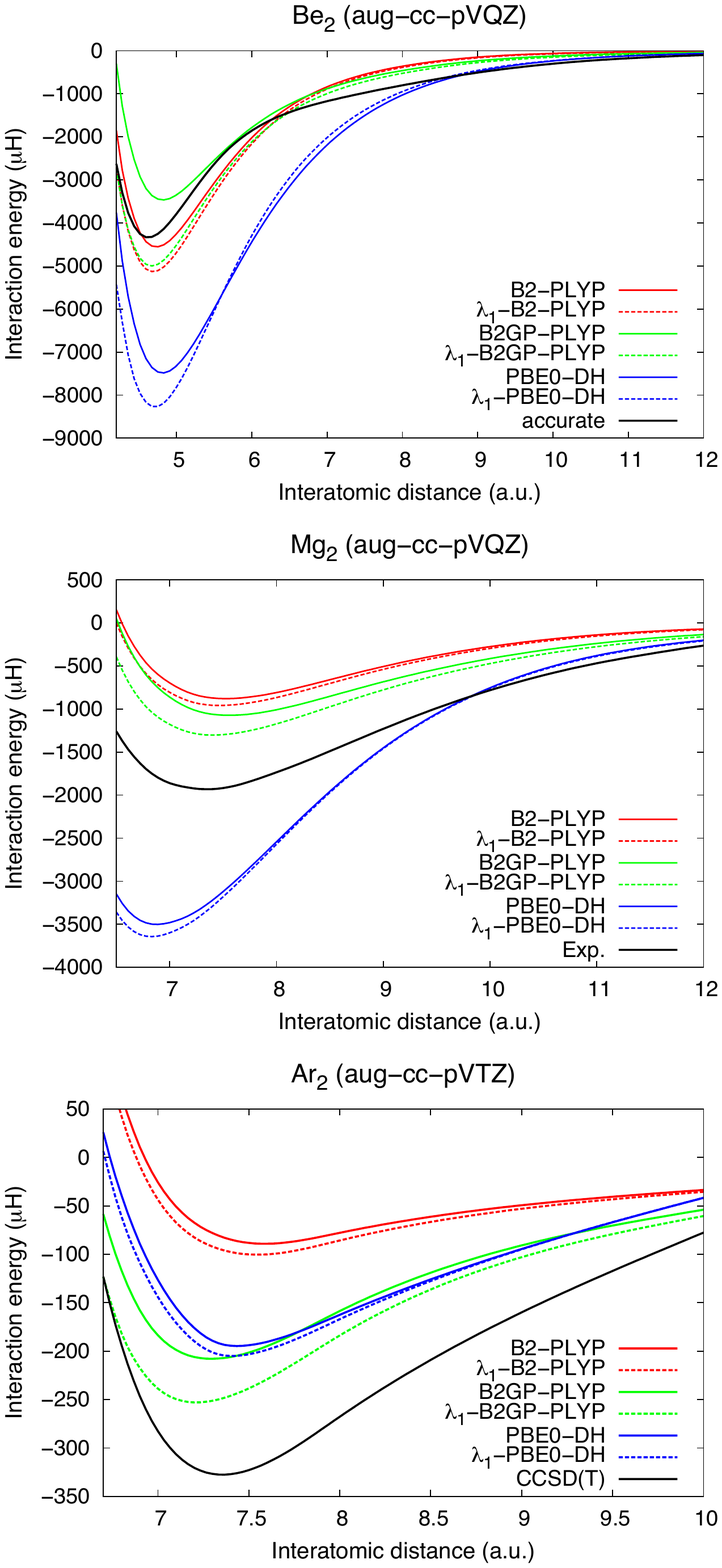}} \\
\end{tabular}
\end{center}
\end{figure}

\begin{description}

\item[Figure \ref{Potcurves_h2_n2}:] Potential curves for H$_2$ (top) and N$_2$ (bottom). The "exact" curves are
taken from Ref.~\cite{exact_pot_curves_h2_n2} 

\item[Figure \ref{HL_gap}:] Single hybrid ($\Delta E_{\rm SH}$) and scaled MP2 ($a_{\rm c}\Delta E_{\rm MP2}$) energy
differences between $\lambda_1$- and conventional B2-PLYP, B2GP-PLYP and
PBE0-DH double hybrids calculated
for H$_2$ with respect to the bond distance (left); HOMO-LUMO gap for
both $\lambda_1$- and conventional B2-PLYP, B2GP-PLYP and                             
PBE0-DH double hybrids calculated for H$_2$
with respect to the bond distance (right).      

\item[Figure \ref{Potcurves_vdw}:] Interaction energy curves for Be$_2$, Mg$_2$ and
Ar$_2$. The accurate~\cite{inge_exgem}, experimental~\cite{balfour_expt}
and CCSD(T) curves are taken as reference. 

\end{description}

\clearpage

\clearpage

\clearpage

\clearpage

\begin{table}
\begin{center}
\begin{tabular}{ccccccccccc}
\hline
\hline
 functional & & $a_{\rm x}=\lambda_2$  & & $a_{\rm x}^2$ & &  $a_{\rm c}$ &
 &   $\lambda_1=a'_{\rm x}$ & & $a'_{\rm c}$\\ 
 \hline
B2-PLYP~\cite{2blehybrids_Grimme}  &  & 0.53 &  & 0.28 & & 0.27& & 0.43 & & 0.19\\
B2T-PLYP~\cite{b2t_Tarno}  &  &  0.60&  &  0.36& & 0.31& &  0.38 & &
0.14\\
mPW2-PLYP~\cite{mPW2_Grimme}  & & 0.55&  & 0.30 & & 0.25& & 0.32 & & 0.10\\
mPW2K-PLYP~\cite{b2t_Tarno}  &  & 0.72 &  &0.52  & & 0.42& & 0.41 & &
0.17\\
B2GP-PLYP~\cite{B2GP_Martin}  & & 0.65&  & 0.42 & & 0.36& & 0.40 & & 0.16\\
B2$\pi$-PLYP~\cite{B2pi_Garcia}  & & 0.602&  & 0.362 & & 0.273 & & 0.303
& & 0.092\\
PBE0-DH~\cite{pbe0-dh_Adamo} ($a_{\rm c}=a^3_{\rm x}$) & & 0.50& & 0.25&
& 0.125& & 0.146 & &
0.021\\
\hline
\hline
\end{tabular}
\caption{Regular ($a_{\rm x}$, $a_{\rm c}$) and modified
($a^{\prime}_{\rm x}$, $a^{\prime}_{\rm c}$)
exchange-correlation coefficients corresponding to     
conventional double hybrids.}
\label{lambda1values}
\end{center}
\end{table}
\end{document}